\documentclass[aps, twocolumn, showpacs]{revtex4}
\begin{document}
\title{Comment on "Cyclotron resonance study of the electron and hole velocity
in graphene monolayers"}
\author { S. C. Tiwari\\
Institute of Natural Philosophy\\
c/o 1 Kusum Kutir Mahamanapuri,Varanasi 221005, India}
\begin{abstract}
In this comment it is pointed out that the electron velocity of the same order
as observed in graphene had been measured in GaAs submicron devices long ago. Particle-
antiparticle asymmetry related with electron and hole effective masses in graphene seems
puzzling as hole in a condensed matter system cannot be treated as anti-electron.
It is argued that there should be a universal electrodynamics for QHE and superconductivity.
In this context attention is drawn to the new approach based on massless electron and the interpretation that magnetic field represents angular momentum of the photon fluid.
Measurement of electron velocity in graphene and GaAs in parallel is suggested for testing
the massless electrodynamics.
\end{abstract}
\maketitle

Sciencewatch (CERN Courier May 2007) highlighted the reported observation of QHE in 
graphene at room temperature. A phrase 'the fact that its electrons act in many ways like 
massless Dirac fermions' attracted my attention as I thought massless electron model proposed
more than two decades ago, and comprehensively discussed in a monograph \cite{1} may
have found an experimental validation in graphene. A careful reading of the paper
\cite{2} shows that 'exceptionally high electron velocities' of the order of $1.117x10^8$
cm/sec in graphene are, in fact, in the range observed in GaAs ballistic transport and
velocity overshoot \cite{3}. The concluding remark in \cite{2} that the band structure 
of monolayer graphene is not well understood at present is quite reasonable. However
the one on particle-antiparticle asymmetry is puzzling. Note that effective
masses of electron and hole differ considerably in general; a hole in condensed matter
system cannot be treated as an antiparticle of electron.

Though it is well known, perhaps it is useful to remind ourselves that Dirac in his
famous work first identified antiparticle of electron to be proton-the only known
positively charged elementary particle at that time. Weyl- a mathematician argued
that symmetry on mathematical grounds demanded the mass of antiparticle equal to 
that of electron-later discovered as positron. Secondly the linear dispersiom
relation by itself, the approximate relation derived by Wallace for graphite, does not
mean that hole could be treated as antiparticle. A massive (complex) Klein-Gordon field
does not have linear relation between energy and momentum yet it has antiparticle, and
photon has a linear relation yet there is no antiphoton (in the standard QFT).

In a more general context I would like to raise the following question: Is there a
universal electrodynamics for QHE and superconductivity? First we note an important
fact: quantized Hall resistance and magnetic flux quantum involve only fundamental
constants (c,h, and e) and do not depend on the material properties or size of the 
samples. There exist compelling arguments discussed in detail in Sec. 9.2 of \cite{1}
indicating a universal electrodynamics assuming massless electron and the
interpretation that magnetic field is angular momentum flux of photon fluid.
The measurement of electron velocity in superconductor and QHE device has been
advocated in the book. That the problem of Cooper pair mass is nontrivial has been
underscored by Mishonov \cite{5} quoting de Gennes remark from Tinkham's book and Ginzburg's private communication. Amongst three characteristics (zero resistance, Meissner effect, and
flux quantization) the explanation of zero resistance in BCS theory seems unsatisfactory.
Feynman says \cite{5} : "There is no resistance because all the electrons are collectively
in the same state. In the ordinary flow of current you knock one electron or the
other out of the regular flow, gradually deteriorating the general momentum. But here
to get one electron away from what all the others are doing is very hard because
of the tendency of all Bose particles to go in the same state. A current once started,
just keeps on going forever." This is hardly an explanation, further persistent
current is not quite the same as zero resistance. Weinberg \cite{6} seems to have
addressed the problem better- he remarks that the absence of resistance (other than
that in closed rings for persistent current) can be understood considering time-dependent
effects. Relating the voltage with the time derivative of a postulated Goldstone
boson and defining a suitable Hamiltonian of the system he offers an explanation for zero resistance. The role of Cooper pairs, however remains obscure.

In a new approach field-free filamentary tubular structures in condensed matter system have 
been envisaged in which electrons move with the velocity of light \cite{1}. Few collisions
could result in quantized average drift velocity. This idea has been used to understand
QHE and superconducting electrodynamics in \cite{1}. A simple model for the ballistic
transport \cite{7} found remarkable support in new experiments at the IBM as noted by
Heiblum \cite{8} :"... new experiments, not yet published, show that ballistic transport 
through GaAs regions wider than previously reported (about 800 A wide), result with... .
These results are qualitatively in agreement with Tiwari's comment".

Recall that the eigenvalue of velocity operator in Dirac's theory is equal to $\pm c$ ;
localizing electron wavepacket in a region smaller than electron's Compton wavelength
leads to a mixing of negative and positive frequency states- the zitterbewegung of
Schroedinger. Point electron theory of Dirac conflicts with the proposed extended 
structure \cite{1,9} and electron motion with light velocity is considered unphysical.
In our model electron is an extended spatio-temporal 2+1 dimensional object, and
has velocity in field-free region (or rather intrinsic) equal to c normal to the
2-spatial plane. Inertia is interpreted as a consequence of collisions or viscosity
of the space, see modified Schroedinger equation derived using this idea in PLA
(1988) \cite{9}.

Obviously testing the radical idea of massless electron would have far reaching
implications on fundamental physics. I suggest time-of-flight technique
used in GaAs \cite{10} be explored for graphene. It seems InAs semiconductor and
Nb superconductor have transparent interface i.e. no Schottky barrier \cite{11}.
In that case long superconductor could be used to investigate the velocity of 
electron/ Cooper pairs. In conclusion, studies on electron transport in GaAs and 
graphene carried out in parallel might give useful new physics. For GaAs properties
we refer to \cite{12}.

 The Library facility at Banaras Hindu University is acknowledged.

\end{document}